\documentstyle[11pt]{article}
\begin{document}
\baselineskip 18pt
\begin{titlepage}
\centerline{\large\bf New Constraint on Weak CP Phase}
\centerline{\large\bf Rephasing Invariant and Maximal CP Violation }
\vspace{1cm}

\centerline{ Yong Liu }
\vspace{0.5cm}
\centerline{ Lab of Numerical Study for Heliospheric Physics (LHP)}
\centerline{  Chinese Academy of Sciences}
\centerline{  P. O. Box 8701, Beijing 100080, P.R.China}
\vspace{0.2cm}
\centerline{  E-mail: yongliu@ns.lhp.ac.cn}
\vspace{2cm}

\centerline{\bf Abstract}

By using the relation between CP-violation phase and the mixing angles
in Cabibbo-Kobayashi-Maskawa matrix postulated by us before, the rephasing
invariant is recalculated. Furthermore, the problem about maximal CP
violation is discussed. We find that the maximal value of
Jarlskog's invariant is about $0.038$. And it presents
at $\alpha\simeq 71.0^0$, $\beta\simeq 90.2^0$ and $\gamma\simeq 18.8^0$
in triangle {\bf db}.
\\\\
PACS number(s): 11.30.Er, 12.10.Ck, 13.25.+m

\vspace{2cm}
\end{titlepage}

\centerline{\large\bf New Constraint on Weak CP Phase}
\centerline{\large\bf Rephasing Invariant and Maximal CP Violation }
\vspace{0.5cm}

Since the discovery of CP-violation in neutral kaon system in 1964 [1],
more than thirty years have passed. Although a great progress has been
made during these years [2-6], such as establishing the phenomenological
structure of the effects, classifying the CP parameters and contructing
the gauge theory models [7-14], our understanding about CP violation is
still very poor. On the experimental side, the kaon system remains
the only place where CP violaion is observed, though the $B$ meson system
is the best probe to it as is widely accepted [15-16]. On the theoretical
side, the origin of CP violation is not so clear, and the correctness of
the standard Cabibbo-Kobayashi-Maskawa mechanism is far from being proved.

In the standard model of three generations, CP violation originates from
the phase angle present in the unitary Cabibbo-Kobayashi-Maskawa matrix.
Mathematically, it is permited that, only one phase angle exists in a three
by three unitary matrix with the exception of three Eulerian angles.
However, the weak CP phase which cannot be eliminated by any means, is
introduced somewhat artificially. The requirement that it has nothing to
do with the three mixing angles is only due to the mathematical rather than
a physical reason. It is naturally to ask whether there is an intrinsic
relation between the phase and the three mixing angles.

Recently, we found that the CP-violation phase and the other three mixing
angles satisfy the following relation [17-18]
\begin{equation}
sin\frac{\delta}{2}=\sqrt{\frac{sin^2\theta_1+sin^2\theta_2+sin^2\theta_3
-2(1-cos\theta_1 cos\theta_2 cos\theta_3)}{2 (1+cos\theta_1) (
1+cos\theta_2) (1+cos\theta_3)}}
\end{equation}
where $\theta_i\; (i=1,\; 2,\; 3)$ are the corresponding angles in the
standard KM parametrization matrix
\begin{equation}
V_{KM}= \left (
\begin{array}{ccc}
   c_1 & -s_1c_3& -s_1s_3 \\
   s_1c_2 & c_1c_2c_3-s_2s_3e^{i\delta}& c_1c_2s_3+s_2c_3e^{i\delta}\\
   s_1s_2 & c_1s_2c_3+c_2s_3e^{i\delta}& c_1s_2s_3-c_2c_3e^{i\delta}
\end{array}
\right )
\end{equation}
with the standard notations $s_i=sin\theta_i$ and $c_i=cos\theta_i$
are used.

The geometry meaning of Eq.(1) is very clear, $\delta$ is the solid
angle enclosed by three angles $\theta_1$, $\theta_2$ and $\theta_3$,
or the area to which the solid angle corresponds on a unit sphere.
It should be noted that, to make $\theta_1$, $\theta_2$ and $\theta_3$
enclos a solid angle, the condition
\begin{equation}
\theta_i+\theta_j > \theta_k  \;\;\;\; (i\neq j \neq k \neq i.
\;\; i,j,k=1,2,3)
\end{equation}
is needed. Now, we find that the CP-violation seems to originate in a
geometry reason.

With the discussion on the "maximal" CP violation in various
parametrizations of the KM matrix [19-22], It is found [23-26] that all
the CP
nonconservation effects are proportional to a universal factor $X_{CP}$
defined as,
\begin{equation}
X_{CP}=s_1^2 c_1 s_2 c_2 s_3 c_3 s_{\delta}.
\end{equation}
$X_{CP}$ is also called the Jarlskog's invariant in the relevant
references.

In fact, only those functions of $V_{KM}$ which are invariant under the
rephasing operation of quark fields can be observable.
From $V_{KM}$ we can construct [27] nine squared moduli invariants
$\Delta_{\alpha \beta}^{(2)}\equiv \mid (V_{KM})_{\alpha \beta}
\mid ^2$ and nine quartic invariant
function
\begin{equation}
\Delta_{\alpha \rho}^{(4)}\equiv (V_{KM})_{\beta \sigma}
(V_{KM})_{\gamma \tau}(V_{KM}^\ast)_{\beta \tau}
(V_{KM}^\ast)_{\gamma \sigma}
\end{equation}
here, the summation convention for repeated
indices is used and $\alpha,\; \beta,\; \gamma\; (\rho,\;
\sigma,\; \tau)$ are taken cyclic.
Owing to the unitary constraint, only four squared-moduli
and one quartic term are independent. In the mean time, all the nine quartic
term have the same imaginary part, it is just $X_{CP}$ shown in Eq.(4).

Substitue Eq.(1) into Eq.(4), we obtain
\begin{equation}
X_{CP}=s_1^2 c_1 s_2 c_2 s_3 c_3
\frac{(1+c_1+c_2+c_3) \sqrt{s_1^2+
s_2^2+s_3^2-2 (1-c_1 c_2 c_3)}}{(1+c_1) (1+c_2) (1+c_3)}
\end{equation}
This is the rephasing invariant after considering the new constraint
on the weak CP phase and the quark mixing angles, based on which, we can
further discuss the maximal CP violation in following.

It is not difficult to find out the maximum of $X_{CP}$, it is
\begin{equation}
X_{CP}^{Max}=0.038296
\end{equation}
and presents at
\begin{equation}
c1=0.49666 \;\;\;\;\;\; c2=c3=0.56766
\end{equation}
In this case, the Wolfenstein parametrization approximate to the third
order of $\lambda$ is invalid. To give a little
sense about the CP violation, we calculate the three angles of the
triangle ($\bf db$) defined as following [28]
\begin{equation}
\alpha=arg\left( -\frac{V_{td} V_{tb}^{*}}{V_{ud} V_{ub}^{*}} \right)
\end{equation}
\begin{equation}
\beta=arg\left(-\frac{V_{cd} V_{cb}^{*}}{V_{td} V_{tb}^{*}} \right)
\end{equation}
\begin{equation}
\gamma=arg\left(-\frac{V_{ud} V_{ub}^{*}}{V_{cd} V_{cb}^{*}} \right)
\end{equation}
From Eq.(1), Eq.(2) and Eq.(8), we get
\begin{equation}
\alpha=71.0^0  \;\;\;\;\;\;
\beta=90.2^0   \;\;\;\;\;\;
\gamma=18.8^0
\end{equation}
It is easy to find that is nearly a right triangle with $0.2^0$
deviation. However, nature has not choose this way of CP violation.
It deviates very far from the present experimental results [29-30].

As a conclusion, we have recalculated the rephasing invariant
with the relation between CP-violation phase and the mixing angles
in CKM matrix postulated by us before being used.
The problem about maximal CP
violation is discussed. We find that the maximal value of
Jarlskog's invariant is about $0.038$. And it presents
at $\alpha\simeq 71.0^0$, $\beta\simeq 90.2^0$ and $\gamma\simeq 18.8^0$
in the triangle {\bf db}.

Based on the above model-independent results, we can extract some
limits on the experiments in $B^0-\overline{B^0}$ and
$D^0-\overline{D^0}$ system etc. The further work will be
reported in the future.


\vspace{0.5cm}

\begin{thebibliography}{30}
\bibitem{1} J.H.Christenson, J.W.Cronin, V.L.Fitch and R.Turlay,
            Phys.Rev.Lett.{\bf 13},138(1964).
\bibitem{2} L.L.Chau, Phys.Rept.{\bf 95},1(1983).
\bibitem{3} E.A.Paschos and U.Turke, Phys.Rept.{\bf 4},145(1989).
\bibitem{4} $CP \;\; Violation$ Ed. L.Wolfenstein, North-Holland,
             Elsevier Science Publishers B.V. 1989.
\bibitem{5} $CP\;\; Violation$ Ed. C.Jarlskog. World Scientific Publishing
             Co.Pte.Ltd 1989.
\bibitem{6} A.Pich, CP Violation, Preprint CERN-TH.7114/93. Dec.1993.
\bibitem{7} J.Prentki and M.Veltman, Phys.Lett.{\bf 15},88(1965).
\bibitem{8} T.D.Lee and L.Wolfenstein, Phys.Rev.B{\bf 138},1490(1965).
\bibitem{9} L.Wolfentein, Phys.Rev.Lett.{\bf 13},562(1964).
\bibitem{10} J.Bernstein, N.Cabibbo and T.D.Lee, Phys.Lett.{\bf 13},
            146(1964).
\bibitem{11} J.S.Bell and J.K.Perring, Phys.Rev.Lett.{\bf 13},348(1964).
\bibitem{12} L.B.Okun', Sov.J.Nucl.Phys.{\bf 1},670(1965).
\bibitem{13} N.Cabibbo, Phys.Rev.Lett.{\bf 10},531(1963).
\bibitem{14} M.Kobayashi and T.Maskawa, Prog.Theor.Phys.{\bf 42},652(1973).
\bibitem{15} A.B.Carter and A.Danda, Phys.Rev.D {\bf 23},1567(1981).
\bibitem{16} I.I.Bigi and A.I.Sanda, Nucl.Phys.B {\bf 193},85(1981).
\bibitem{17} J.L.Chen, M.L.Ge, X.Q.Li and Y.Liu, New Viewpoint to the
             Source of Weak CP Phase, Preprint hep-ph/9711330.
\bibitem{18} Yong Liu and Jing-Ling Chen, New Constraint on the Parameters
             in Cabibbo-Kobayashi-Maskawa Matrix of Wolfenstein's
             Parametrization, Preprint hep-ph/9711293.
\bibitem{19} O.W.Greenberg, Phys.Rev.D,{\bf 32},1841(1985).
\bibitem{20} M.Gronau and J.Schechter, Phys.Rev.Lett.,{\bf 54},385(1985).
\bibitem{21} L.Wolfenstein, Phys.Lett.B {\bf 144},425(1984).
\bibitem{22} D.Hochberg and R.G.Sachs, Phys.Rev.D,{\bf 27},606(1983).
\bibitem{23} L-L.Chau and W-Y.Keung, Phys.Rev.Lett.,{\bf 53},1802(1984).
\bibitem{24} F.J.Botella And L-L.Chau, Phys.Lett.B,{\bf 168},97(1986).
\bibitem{25} C.Jarlskog, Phys.Rev.Lett. {\bf 55},1039(1985).
\bibitem{26} C.Jarlskog, Z.Phys.C,{\bf 29},491(1985).
\bibitem{27} J.F.Donoghue, E.Golowich and B.R.Holstein, $Dynamics\;\;of\;\;
              the\;\;Standard\;\;Model$ Cambridge University Press, 1992.
              P.60$\sim$P.69.
\bibitem{28} V.Gibson, J.Phys.G, {\bf 23}, 605(1997).
\bibitem{29} A.J.Buras, {\sl Proc. 28th Int. Conf. on High Energy Physics
             (Warsar)} TUM-HEP-259/96.
\bibitem{30} A. Ali and D. London, DESY 96-140 (1996).
\end{thebibliography}
\end{document}